\documentclass[aps,preprint,nofootinbib]{revtex4}
\pdfoutput=1
\usepackage{comment}
\usepackage{graphics}
\usepackage{slashed}
\usepackage{amssymb}
\usepackage{lscape}
\usepackage{amsmath}
\usepackage{rotating}
\usepackage{graphicx}
\usepackage{amsthm}
\usepackage{MnSymbol}
\usepackage{hyperref,color,xcolor}
\usepackage{tensor}
\usepackage{bm}
\usepackage{hyperref}
\graphicspath{ {images/} }
\theoremstyle{plain}
\newtheorem{theo}{Theorem}

\theoremstyle{definition}
\newtheorem*{defn*}{Definition}

\begin{document}
\title{Particle-antiparticle duality from an extra timelike dimension}
\author{$^{2}$ Marcos R. A. Arcod\'{\i}a\footnote{E-mail address: marcodia@mdp.edu.ar},  $^{1,2}$ Mauricio Bellini
\footnote{E-mail address: mbellini@mdp.edu.ar} }
\address{$^1$ Departamento de F\'isica, Facultad de Ciencias Exactas y
Naturales, Universidad Nacional de Mar del Plata, Funes 3350, C.P.
7600, Mar del Plata, Argentina.\\
$^2$ Instituto de Investigaciones F\'{\i}sicas de Mar del Plata (IFIMAR), \\
Consejo Nacional de Investigaciones Cient\'ificas y T\'ecnicas
(CONICET), Mar del Plata, Argentina.}
\begin{small}
\begin{abstract}
It is a well known fact that the usual complex structure on the real Clifford Algebra (CA) of Minkowski spacetime can be obtained by adding an extra time-like dimension, instead of the usual complexification of the algebra. In this article we explore the consequences of this approach and reinterpret known results in this new context.

We observe that Dirac particles and antiparticles at rest can be interpreted as eigenstates of the generator of rotations in the plane formed by the two time-like coordinates and find an effective finite scale for the extra dimension when no EM fields are present (without postulating compactness). In the case of non-vanishing EM fields, we find a gauge condition to preserve such a scale.
\end{abstract}
\end{small}
\maketitle

\section{INTRODUCTION}

Since the theory of Kaluza-Klein was proposed \cite{KK}, the idea of a five dimensional spacetime has been widely explored in physics. The Kaluza-Klein theory allows the unification of the Einstein field equations and Maxwell electromagnetism by considering a 5D manifold with a compact fifth coordinate.

In the early nineties Wesson along with collaborators proposed the existence of a non compact extra dimension and presents the theory of space-time-matter (STM) or induced matter theory (IMT)\cite{STM}, as a way to induce 4D Einstein's equations from a 5D Ricci-flat manifold, making use of the Campbell-Magaard theorem.

Following the ideas of IMT, applications to particle physics and quantum mechanics have also been considered\cite{ParticleQM}. In these regard, physical quantities (charge, mass) are obtained as geometric parameters from a higher-dimensional spacetime. Since one is adding an extra dimension to the spacetime, the nature of it (space-like or time-like) must be taken into account. Spacetimes with time-like extra dimension (generally referred to as two-times spacetimes, e.g. \cite{2times}), and space-like extra dimension (e.g. \cite{Sanchez}), has been considered with different applications in cosmology and particle physics.

In this article we are interested in analyzing the Dirac theory and the consequences of an extra time-like dimension in spacetime from a Clifford algebra point of view. We shall also provide an interpretation for what we call the time plane and relate it to the particle/antiparticle character of a Dirac spinor field. 

Theories with more than two time-like dimensions has been proposed \cite{Bars}, and also plenty of work has been done on spaces of signature $(+---+)$ in the context of anti de Sitter spaces \cite{AdS}. 

Previous studies has been made in 5D spinors \cite{Vignolo, Sanchez, Ma, Kocin}, but in this article we focus more on the nature of the extra dimension and the structure of the Clifford algebra without studying gravitational cases. A central part of the Dirac theory, which sometimes is not thoroughly studied, is the Clifford algebra (or geometric algebra) of spacetime\cite{Lounesto, VdR, Gallier}. Starting from this algebra, the spinors are defined as representation spaces for it, and then we introduce the Dirac equation.

Interestingly, an extra dimension naturally emerges on the Dirac theory if we take into account the following fact: the ordinary Minkowski complex Clifford algebra can be obtained as a real algebra with an extra time-like dimension \cite{Lounesto, VdR}, as will be detailed in the article. This is a very well known fact in the theory of Clifford algebras. Indeed, we can not distinguish (except for grading) between the complexified 4D Minkowski Clifford Algebra and the real 5D Clifford algebra: they are isomorphic.

In this work we consider the existence of a time-like extra dimension, we analyze its Clifford algebra and Spin group. Then, we obtain the related massive 4D Dirac equation (with and without an electromagnetic potential) from a massless 5D Dirac equation. We observe that the particles and antiparticles rest solutions for the Dirac field, can be seen as eigenvectors for the generator of a rotation in the defined time plane. We also conclude that, when an electromagnetic field is present, in addition to the time plane rotation we also have a gauge transformation on the spinor. 

In this article we don't ask for the extra dimension to be compact, but, when no electromagnetic fields are present, we obtain an effective scale for this dimension. Additionally, in the presence of electromagnetic fields we obtain a gauge condition to keep the aforementioned scale.

\section{Clifford Algebras preliminaries}

\subsection{Definitions and general results}\label{CAgeneral}

Given a real n-dimensional vector space $V$ with a bilinear symmetric form $\varphi:{V}\times{V}\rightarrow{\mathbb{R}}$, we say that $\Phi:{V}\rightarrow{\mathbb{R}}$, defined by $\Phi(v)=\varphi(v,v)$, is the associated \emph{quadratic form} and we call the pair $(V,\Phi)$ a quadratic space. If the form $\varphi$ is non-degenerate we say the quadratic space is regular. Since $\varphi(a,b)=\frac{\Phi(a+b)-\Phi(a)-\Phi(b)}{2}$, one doesn't lost information when passing from $\varphi$ to $\Phi$.

An important subgroup of the group of linear isomorphisms on this vector space is the \emph{group of isometries} that we shall denote by $\text{O}(\Phi)$, and define by \[\text{O}(\Phi)=\{f\in\text{Aut}{(V)}:\Phi(f(v))=\Phi(v) \ \ \forall{v}\in{V}\}.\]
An important subgroup of this group is that composed by positive determinant transformations, $\text{SO}(\Phi)$:
\[\text{SO}(\Phi)=\{f\in\text{O}(\Phi): \det(f)>0\}.\]
For any quadratic space, one can build an associative unitary real algebra $\text{Cl}(\Phi)$, called the \emph{Clifford algebra (CA) for $(V,\Phi)$}. It is possible to define that algebra in different equivalent ways. Here, we shall do it as follows: Let $\{e_{1},..,e_{n}\}$ be a basis for the vector space $V$, and $\varphi_{ij}$ the matrix elements of the bilinear form $\varphi$ in the given basis. The CA is defined by the generators $\{E_{1},...,E_{n}\}$, with the relations:
\begin{equation}
E_{i}E_{j}+E_{j}E_{i}=2\varphi_{ij}\bm{1},
\end{equation}
where $\bm{1}$ is the unit in $\text{Cl}(\Phi)$. We shall just mention that this algebra can also be constructed as a quotient algebra of the tensor algebra modulo certain ideal\cite{Gallier}, but we shall not go further in this subject. Since the tensor algebra is generally considered a real algebra, the CA also happens to be a real algebra. Because there is an injective function from $V$ to the CA, via $e_{i}\mapsto{E}_{i}$, by abuse of notation we shall refer to the generators of this algebra as $e_{i}$. In the same way, we shall refer to the subspace $\text{span}_{\mathbb{R}}\{E_{1},...,E_{n}\}\subseteq\text{Cl}(\Phi)$, as $V$. It happens that this algebra is finite dimensional with dimension $2^{n}$ \cite{Lounesto}.

We have that a basis for this algebra is the set:
\begin{equation}
e_{i_{1}}...e_{i_{k}} \ : \ 1\leq{i_{1}}<i_{2}<...<i_{k}\leq{n}.
\end{equation}
Using this fact, we will say that an element of the form $\sum_{i<{j}}A^{ij}e_{i}e_{j}$ is a \emph{bivector} or \emph{2-vector} and in general a \emph{k-vector} is an element of the form: \[ \sum_{i_{1}<...<i_{k}}A^{i_{1}...i_{k}}e_{i_{1}}...e_{i_{k}},\]
and a \emph{0-vector} is an element proportional to $\bm{1}$. We call ${\bigwedge}_{k}(V)$ or just ${\bigwedge}_{k}$ the vector space of $k-vectors$, which means ${\bigwedge}_{1}=V$. It can be seen that $\dim({\bigwedge}_{k})={n \choose k}$. Hence there is just one independent $n$-vector and also one independent $0$-vector for every quadratic space. The unitary n-vector is also called \emph{the pseudoscalar} of the algebra.

We define the \emph{grade involution} $\alpha$ on a basis element  $e_{i_{1}}...e_{i_{k}}$ as: $\alpha({e_{i_{1}}...e_{i_{k}}})=(-1)^{k}e_{i_{1}}...e_{i_{k}}$, and extend it to any element as an algebra homomorphism on $\text{Cl}(\Phi)$.

It can be proven that this morphism induces a $\mathbb{Z}_{2}$ grading in the algebra, splitting it into $\text{Cl}(\Phi)=\text{Cl}(\Phi)^{0}\oplus\text{Cl}(\Phi)^{1}$, where $\text{Cl}(\Phi)^{i}=\{x\in\text{Cl}(\Phi):\alpha(x)=(-1)^{i}x\}$. It is true that $\text{Cl}(\Phi)^{0}$ is a subalgebra of $\text{Cl}(\Phi)$ while $\text{Cl}(\Phi)^{1}$ is not. Furthermore, it happens that \[\text{Cl}(\Phi)^{0}=\bigoplus_{i \ even}{\bigwedge}_{i}(V) \quad \text{and} \quad \text{Cl}(\Phi)^{1}=\bigoplus_{i \ odd}{\bigwedge}_{i}(V) . \]
We define \emph{reversion}, $t$, on a basis element $e_{i_{1}}...e_{1_{k}}$ as $t(e_{i_{1}}...e_{i_{k}})=e_{i_{k}}...e_{i_{1}}$, and extend it as an algebra anti-morphism (meaning $t(a.b)=t(b).t(a)$, details can be found in \cite{Gallier, VdR}).

Using the previous functions we define the \emph{Clifford conjugation} on any CA element $x$ as $\overline{x}=(t\circ\alpha)(x)=(\alpha\circ{t})(x)$. This is an algebra anti-morphism and in a basis element $e_{i_{1}}...e_{i_{k}}$ it can be seen to be $\overline{e_{i_{1}}...e_{i_{k}}}=(-1)^{k}e_{i_{k}}...e_{i_{1}}$.

Last, we define the \emph{norm} $N(x)$ of an element $x$ in the CA as $N(x):=x\overline{x}=\overline{x}{x}$. An important feature of this function is that $N(v)=-\Phi(v)\bm{1}$ for any $v\in{V}$.

Although the algebra is constructed as a real algebra, from a real quadratic space, in occasions we need to work with the \emph{complexified} version of the algebra. Roughly speaking complexifying the algebra consists in allowing scalars in the linear combinations to be complex numbers, transforming an $\mathbb{R}$-algebra into a $\mathbb{C}$-algebra. Mathematically this is attained by building the algebra $\text{Cl}_{\mathbb{C}}(\Phi):=\mathbb{C}\otimes\text{Cl}(\Phi)$ with the product trivially defined.

It can be shown that any regular quadratic space admits an orthogonal basis, in the following sense: For every regular quadratic space there's a basis $\{e_{1},...,e_{p},e_{p+1},...,e_{p+q}\}$ such that:
\begin{equation}\label{signaturas}
\varphi(e_{i},e_{j})=
\begin{cases}
0 &  i\neq{j} \\
1 &  i=j\in\{1,...,p\} \\
-1 &  i=j\in\{p+1,...,p+q\}
\end{cases},
\end{equation}
with $p+q=n$. We say that $(p,q)$ or $\{\underbrace{+,+,...,+}_{p-times},\underbrace{-,-,...,-}_{q-times}\}$ is the signature of $\varphi$.\\
We will denote the quadratic form of signature $(p,q)$ as $\Phi_{p,q}$ and its CA as $\text{Cl}_{p,q}$.

\subsection{The Clifford-Lipschitz group and its Lie algebra}\label{CLgroup}

In the CA there are certain groups with special properties. All of them are subgroups of the group of units, $\text{Cl}(\Phi)^{*}$, of the given Clifford Algebra. These groups are closely related to the isometries of the quadratic space and to the well known adjoint action of the CA.
In this section we will briefly define the Clifford-Lipschitz group and also give its Lie algebra without proof. Detailed calculations can be found in \cite{VdR}.
Let $x\in\text{Cl}(\Phi)^{*}$, then there exists $x^{-1}$ such that $\bm{1}=xx^{-1}=x^{-1}x$. The Clifford-Lipschitz group $\Gamma(\Phi)$ is defined as follows:
\begin{equation}
\Gamma(\Phi)=\{x\in\text{Cl}(\Phi)^{*}: xvx^{-1}\in{V} \ \forall \ v\in{V}\}.
\end{equation}
Here we are using the injection $V\hookrightarrow\text{Cl}(\Phi)$ from section \ref{CAgeneral} for $xvx^{-1}$ to be well defined.

As it is explained in \cite{VdR} the Lie algebra of this Lie group is the set:
\begin{equation}
\gamma(\Phi)={\bigwedge}_{2}(V)\oplus{Z(\text{Cl}(\Phi))} \ ,
\end{equation}
with $Z(\text{Cl}(\Phi))$ the center of the CA:
\begin{equation}
Z(\text{Cl}(\Phi))=\{x\in\text{Cl}(\Phi) \ :\ xa=ax \ \forall{a}\in\text{Cl}(\Phi)\},
\end{equation}
and the Lie algebra bracket being the commutator in the CA, $[x,y]=xy-yx$.

In particular, we have $Z(\text{Cl}_{1,3})=\mathbb{R}\bm{1}=\bigwedge_{0}$ and $Z(\text{Cl}_{2,3})=\mathbb{R}\bm{1}\oplus\mathbb{R}i\cong{\mathbb{C}}$, with $i=e_{0}e_{1}e_{2}e_{3}e_{4}$, the pseudoscalar of $\text{Cl}_{2,3}$.

As in the previous section, we will call $\Gamma_{p,q}$ the Clifford-Lipschitz group of the bilinear forms considered in \ref{signaturas}.

Since we are dealing with Lie groups and given the discrepancies between the physics and mathematics literature, it is a good point to state what definition we are going to use for a \emph{generator} of an element $\bm{S}$ in a Lie group $G$.

Recall that if $G$ is a Lie group, then it is a manifold, and its Lie algebra $\mathfrak{g}$ is defined as the tangent space of that manifold at the point $\bm{1}\in{G}$. Given an element $\bm{S}\in{G}$ connected to the identity, there exists a path $\beta:[0,1]\rightarrow\text{G}$, with $\beta(0)=\bm{1}$ and $\beta(1)=\bm{S}$. We say that $\frac{d\beta(t)}{dt}|_{t=0}$ is \emph{the generator of $\bm{S}$}. Observe that the generators of elements in $G$ belong to $\mathfrak{g}$ the Lie algebra of $G$. It can be seen that this is a good definition.

\subsection{The Spin group and its Lie algebra}

The \emph{Spin group} of a certain CA, $\text{Spin}(\Phi)$, is a subgroup of the Clifford-Lipschitz group. This group is defined as follows:
\begin{equation}
\text{Spin}(\Phi)=\{x\in\text{Cl}(\Phi)^{0}\ | \ xvx^{-1}\in{V} \ \forall{v}\in{V} ; \ N(x)\in\{-1,+1\}\}.
\end{equation}
We will call $\text{Spin}(p,q)$ the Spin groups of the bilinear forms defined in \ref{signaturas}.

An important property of this group is that it is a double cover of the isometry group for the quadratic form $\Phi$. This is stated in the following theorem:
\begin{theo}\label{teoAd}
There exists a surjective group homomorpism $\text{Ad}:\text{Spin}(\Phi)\rightarrow\text{SO}(\Phi)$ with $\ker(\text{Ad})=\{-1,+1\}$. The mapping Ad is given by: \[\text{Ad}(\bm{S})(v)=\bm{S}v\bm{S}^{-1} \ ,\] recall that we use the injection in section \ref{CAgeneral} in order to have $V\subseteq{\text{Cl}(\Phi)}$.
\end{theo}

This surjective group homomorphism can be extended to the Clifford-Lipschitz group:
\begin{equation}
\begin{gathered}
\text{Ad}:\Gamma(\Phi)\rightarrow\text{O}(\Phi) \quad if \ \dim(V) \text{ is even, with }\ker{\text{Ad}}=\mathbb{R}^{*}\bm{1} \\
\text{Ad}:\Gamma(\Phi)\rightarrow\text{SO}(\Phi) \quad if \ \dim(V) \text{ is odd, with }\ker{\text{Ad}}=\mathbb{R}^{*}\bm{1}\oplus\mathbb{R}^{*}\bm{I}, \ \text{where }\ \bm{I}=e_{1}...e_{n}
\end{gathered}
\end{equation}

The function $\text{Ad}$ is indeed a Lie group representation. $\text{Ad}$ is called the \emph{adjoint representation} or \emph{adjoint action} of Spin$(\Phi)$ (or the Clifford-Lipschitz group).

It is well known that any real CA is isomoprhic to a finite-dimensional matrix algebra over $\mathbb{R}, \ \mathbb{C}$ or $\mathbb{H}$ (or direct sum of these algebras), while any complex CA is isomoprphic to a finite-dimensional matrix algebra over $\mathbb{C}$ (or direct sum of these algebras) (see \cite{Lounesto,VdR}).

In the framework of the Dirac theory one works with the matrix form of the CA, and with what is called the \emph{regular representation} of the CA (which induces a representation of the Spin group). In this case the representation space is the space of \emph{spinors} (actually algebraic spinors), which are the column vectors for which the CA matrices represent a linear transformation. For instance, if $\text{Cl}(\Phi)\cong\mathcal{M}(2,\mathbb{C})$ then, a spinor is an element in $\mathbb{C}^{2}$.

It happens that the space of spinors can be identified with a minimal left ideal in the CA. We will not enter in the details of this construction (see \cite{VdR,Lounesto}) but this is a very well known fact in the CA theory, and indeed it would be a more elegant way to introduce spinors; however, for sake of simplicity, we content ourselves with the definition given above.

Let $\Psi$ be a spinor, then an element in $A\in\text{Cl}(\Phi)$ acts on $\Psi$ just by left multiplication using the matrix representation stated above, $\Psi\mapsto{A\Psi}$.

The theorem \ref{teoAd} is an important result because for every isometry of the vector space $V$ it provides an element (two, actually) in the Spin group. If we further require the element in the Spin group to be connected to the identity, we get uniqueness. In this sense, the morphism $\text{Ad}$ tells us how does a spinor change when we perform certain coordinate transformation in $\text{SO}(\Phi)$.

Reciprocally, if we decide to make a transformation in the space of spinors $\Psi\mapsto{\bm{S}\Phi}$, with $\bm{S}\in\text{Spin}$ then, $\text{Ad}$ gives us the corresponding transformation induced by it in $V$.

As will be clarified in the following sections, the vector space $V$ will model the coordinate spacetime, while the space of spinors corresponds to a spin $1/2$ physical field. Hence, the relation between the adjoint representation and the regular one reflects the relationship between coordinate space-time and internal space for the configuration of the field. In this sense the elements in the kernel of Ad are somehow an internal symmetry of the theory of spinors.

It can be seen that the Lie algebra $\text{spin}(\Phi)$ of this Lie group is given by:
\begin{equation}
\text{spin}(\Phi)={\bigwedge}_{2}(V),
\end{equation}
and that by virtue of the previous theorem, the following Lie algebra isomorphism holds: $\text{spin}(\Phi)\cong{\text{so}(\Phi)}$.

\section{Clifford algebras in physics and applications to the Dirac theory}

\subsection{The Minkowski real CA and its complexification from an extra dimension}\label{realcomplex}

As stated in the previous sections, the CA is associated to a particular quadratic space. In the realm of physics we encounter this kind of spaces in different contexts. For instance in Newtonian mechanics, space is modeled as the quadratic space $\mathbb{R}^{3}$ with the Euclidean quadratic form (which happens to be a norm), however, the case we will pay more attention to in this article is that of special relativity. In this theory space-time is modeled as a Lorentzian 4-dimensional space. This is, the vector space $\mathbb{R}^{4}$, with coordinates $(x_{0},x_{1},x_{2},x_{3})$ (which we will generically refer to as $x_\mu$) together with the Minkowski bilinear form, whose components in the Cartesian basis are:
\begin{equation}
\eta_{\mu\nu}=
  \begin{cases}
    1 &  \ \mu=\nu=0 \\
   -1 & \ \mu=\nu=j ; \ j \in \{1,2,3\}\\
    0 & \mu\neq\nu
  \end{cases}.
\end{equation}

It is said that we adopt the signature $\{+---\}$ or $(+---)$ or $1,3$. We will call this quadratic regular space \emph{the Minkowski space-time}, and refer to it as $\mathbb{R}^{1,3}$. Thus, the real CA corresponding to this spacetime, $\text{Cl}_{1,3}(\mathbb{R})$, will be the algebra generated by elements $\{e_{\mu}:\mu\in\{0,...,3\}\}$ with the relation:
\begin{equation}\label{CArelation}
e_{\mu}e_{\nu}+e_{\nu}e_{\mu}=2\eta_{\mu\nu}\bm{1},
\end{equation}

According to the well known classification of CA, the real algebra for the Minkowski spacetime with this signature is isomorphic to the algebra of ${2}\times{2}$ matrices with entries in the quaternions.

In this article we are going to work with the Dirac theory of spinors. This theory is derived partially from quantum mechanics, which postulates the existence of a complex Hilbert space of physical states, hence we need to use the complex CA, $\text{Cl}_{1,3}(\mathbb{C})$. This algebra is well known to be isomorphic to the complex algebra of ${4}\times{4}$ matrices with complex entries, $\mathcal{M}(\mathbb{C},4)$. There are infinite matrix representations for the generators as matrices in $\mathcal{M}(\mathbb{C},4)$, but the more popular are perhaps the Dirac and Weyl representations.

We will pay special attention to the following known fact: the CA of $\mathbb{R}^{1,3}$ can be ``complexified'' in an alternative way \cite{Lounesto}, which allow us to keep working with real Clifford algebras. The complexification is accomplished by adding an extra time-like dimension, $x^{4}$, to the Minkowski spacetime and taking the real CA of the 5D spacetime with signature $\{+---+\}$, $\mathbb{R}^{2,3}$. This is possible because the following isomorphisms hold:
\begin{equation}\label{iso}
\text{Cl}_{2,3}(\mathbb{R})\cong\mathcal{M}(\mathbb{C},4)\cong\text{Cl}_{1,3}(\mathbb{C}).
\end{equation}
In this case, the imaginary unit in $\text{Cl}_{1,3}(\mathbb{C})$ is identified with the pseudoscalar $e_{0}e_{1}e_{2}e_{3}e_{4}$ in the 5D real algebra. This identification works because the pseudoscalar squares to $-\bm{1}$, and it also lies in the center of the algebra, $Z\text{(Cl}_{2,3})$. Thus, we have that $Z(\text{Cl}_{2,3})\cong\mathbb{C}$.

It is important to be careful when working in $\text{Cl}_{2,3}(\mathbb{R})$, since we shall use the name $i$ to refer to the pseudoscalar, but we are by no means complexifying $\text{Cl}_{2,3}(\mathbb{R})$. Reciprocally, when we complexify the theory in the usual way, this fifth dimension emerges naturally as the matrix element $\gamma_{5}$, which squares to $1$ and is associated with the chirality of the Dirac spinor fields. Since the isomorphism \ref{iso} holds, the representations of $\text{Cl}_{1,3}(\mathbb{C})$ and $\text{Cl}_{2,3}(\mathbb{R})$ are equivalent, and hence the spaces of spinors are isomorphic.

In this article we shall explore the consequences of considering this extra dimension as a real physical one, particularly in the subject related to the particle/antiparticle interpretation of the spinor field. Following the spirit of the theory of Induced Matter \cite{Wesson}, we will obtain the massive 4D Dirac equation from a massless 5D Dirac equation.

In what follows we will name the 5D coordinates $x_{A}$, and the 5D Minkowski metric $\eta_{AB}=\text{diag}(1,-1-1-1,1)$, hence uppercase latin scripts range from $0$ to $4$. The 4D coordinates will be named $x_{\mu}$ and the 4D metric $\eta_{\mu\nu}=\text{diag}(1,-1-1-1)$, hence lowercase greek scripts range from $0$ to $3$. We shall use latin lowercase scripts (e.g. $i,j,k$), that take the values $1,2,3$.

Although \ref{iso} establishes that the algebras $\text{Cl}_{2,3}(\mathbb{R})$ and $\text{Cl}_{1,3}(\mathbb{C})$ are isomorphic, the isomorphism is not univocally determined. Hence we have to pick an isomorphism that will allows us to do the calculations and provide physical interpretations.

If we define the tilded elements:
\begin{equation}
\tilde{e_{\mu}}=-ie_{4}e_{\mu}=-e_{0}e_{1}e_{2}e_{3}e_{\mu}
\end{equation}
for $\mu\in\{0,...,3\}$, then we have the following:
\begin{equation}
\tilde{e}_{0}=e_{1}e_{2}e_{3}\ ; \ \tilde{e}_{1}=e_{0}e_{2}e_{3}\ ; \ \tilde{e}_{2}=e_{1}e_{0}e_{3}\ ; \ \tilde{e}_{3}=e_{0}e_{1}e_{2} \ .
\end{equation}
Using this, we can see that the algebra generated by $\{\tilde{e}_{0},\tilde{e}_{1},\tilde{e}_{2},\tilde{e}_{3}\}$ is a subalgebra of $\text{Cl}_{2,3}$, and is equal to the algebra generated by $\{{e}_{0},{e}_{1},{e}_{2},{e}_{3}\}$.
\begin{defn*}
We will call a vector $\tilde{V}^{\mu}\tilde{e}_{\mu}$, a \emph{tilded 4-vector}, in opposition to an ordinary 5-vector $V^{A}e_{A}$.
\end{defn*}
An interesting feature of this treatment is that the elements $i\tilde{e_{\mu}}$ now belong to the vector space ${\bigwedge}_{2}(\mathbb{R}^{2,3})$, which is the Lie algebra of the group $\text{Spin}(2,3)$ and hence, the vectors generating the algebra can also be seen as generators of certain coordinate transformations preserving the $5D$ metric tensor in the Spin group. Recall that, for instance, $e_{0}e_{1}$ is the generator of a boost in the direction $x_{1}$ and $e_{1}e_{2}$ is a generator of a rotation in the plane $x_{1}-x_{2}$. If we think of transformations in a space of functions in $\mathbb{R}^{n}$, then $\frac{\partial}{\partial{x_{k}}}$ is the generator of a translation in the $k$ direction.

Since $i\tilde{e_{\mu}}=e_{4}e_{\mu}\in{\bigwedge}_{2}(\mathbb{R}^{2,3})$, $i\tilde{e}_{0}$ is the generator of a rotation in the plane of the two time coordinates, and the elements $i\tilde{e_{i}}\in{\bigwedge}_{2}(\mathbb{R}^{2,3})$, are the generators of Lorentz boosts with respect to the extra time and in the spatial direction $x_{i}$. Furthermore, it can be easily seen that $\tilde{e}_{\mu}\tilde{e}_{\nu}=e_{\mu}e_{\nu}$, and hence rather $[\tilde{e}_{\mu},\tilde{e}_{\nu}]$ or $[e_{\mu},e_{\nu}]$, can be considered as a generator or a Lorentz transformation in the 4D space $(x_{0},x_{1},x_{2},x_{3})$. Also, if we consider a Lorentz transformation in the 4D spacetime $(x_{0},...,x_{3})$ with $\nobreak{x'_{\mu}=\tensor{\Lambda}{^\nu_\mu}x_{\nu}}$, we know that there exists an element $\bm{S}\in\text{Cl}$, such that $\bm{S}{e}_{\mu}\bm{S}^{-1}=\tensor{\Lambda}{^\nu_\mu}e_{\nu}$. Now, if we do the same computation on the $\tilde{e}_{\mu}$ we have the same relation, namely, $\bm{S}\tilde{e}_{\mu}\bm{S}^{-1}=\tensor{\Lambda}{^\nu_\mu}\tilde{e}_{\nu}$. This is due to the fact that $\bm{S}=\exp{(\tensor{\varepsilon}{^{\alpha\beta}}e_{\alpha}e_{\beta})}$, with $\tensor{\varepsilon}{^{\alpha\beta}}$ skew-symmetric. Thus, $\bm{S}$ commutes with $e_{4}$ in $\tilde{e}_{\mu}=-ie_{4}e_{\mu}$, and we arrive to the stated conclusion. This is important because it implies that, when restricted to the 4D space-time, a 4-vector $V^{\mu}e_{\mu}$ transforms in the exact same way as the element $V^{\mu}\tilde{e}_{\mu}$.
Note that since $i\tilde{e}_{0}=e_{4}e_{0}$ commutes with $e_{1}e_{2}$, we can have a spinor that is simultaneously a spin eigenvector and an eigenvector for the 2-time plane rotation generator $e_{4}e_{0}$.

Given a general spinor field $\psi$ we can define its \emph{bilinear covariants}\cite{Lounesto,VdR}, which are quantities that transform as tensors under an isommetry of the quadratic space. Bilinear covariants of order $k$ with $0\leq{k}\leq{n}$ can be defined in an $n$-dimensional vector space, for instance, in 4D Minkowski spacetime we define the scalar, vector, bivector, axial vector and pseudoscalar covariants:
\begin{equation}\label{bilcov}
\sigma=\bar{\psi}\psi,\;\;J^{\mu}=\bar{\psi}\gamma ^{\mu }\psi,\;\;S^{\mu\nu}=\frac{i}{2}\bar{\psi}[\gamma^{\mu},\gamma^{\nu }]\psi,\;\; K^{\mu }=\bar{\psi}\gamma^{5}\gamma^{\mu }\psi,\;\;\omega=i\bar{\psi}\gamma^{5}\psi,
\end{equation}
where $\gamma^{5}=i\gamma^{0}\gamma^{1}\gamma^{2}\gamma^{3}$. It is known that in Minkowski spacetime this quantities determine the spinor field up to a complex factor\cite{Lounesto}. Note that this are sixteen real numbers while the real components of a spinor are only eight, hence they can not be all independent. Indeed they are related by the Fierz-Pauli-Kofink identities\cite{Lounesto,Fabbri}:
\begin{subequations}
\begin{eqnarray}\label{fifi1}
-\omega{S}_{\mu\nu}+\sigma\epsilon_{\mu\nu}^{\;\;\;\alpha\beta}S_{\alpha\beta}&=&\epsilon_{\mu\nu\alpha\beta}J^\alpha {K}^\beta,\\
J_\mu J^\mu+K_\mu K^\mu&=&0=J_\mu K^\mu,\label{fifi2}\\
J_\mu J^\mu&=&\omega^{2}+\sigma^{2}\,.\label{fifi3}  
\end{eqnarray}\label{fifi}
\end{subequations}
Being able to express a spinor in terms of tensor quantities allows us to reduce their degrees of freedom by means of Lorentz transformations on the base vector space. For instance, provided that vector covariant $J^{\mu}$ in equation \ref{bilcov} is time-like, we can perform a Lorentz boost on the spinor to remove the spacial components of $J^{\mu}$. Also, once the boost is made, the axial vector transform to a pure spatial vector (due to \ref{fifi2}), and we can perform a rotation to align the axial vector components to one of the axis of our frame, say the z-axis. Hence, the components of the spinor are reduced from eight to two, by removing six degrees of freedom (three from the boost and three from the rotation). It can be shown that regular spinors (which implies $J^{\mu}$ is time-like) in four dimensions can be written in the chiral basis in the form\cite{Vignolo,Fabbri}:
\begin{equation}\label{4ddof}
\psi=\phi\begin{pmatrix}
		e^{i\beta/2}\\
		0 \\
		e^{-i\beta/2} \\
		0 \\
\end{pmatrix}.
\end{equation}
Thus, although a spinor in four dimensions (as well as in five) have eight real components, not all of them are genuine degrees of freedom, since they can be removed by Lorentz transformations. As it can be seen from \ref{4ddof}, the genuine degrees of freedom of a regular spinor in four dimensions are only two.

Let's note that in five dimensions the set of bilinear covariants amount to 32 quantities while they are only 16 in the 4D spacetime, however, since the spinor is determined up to a complex factor by the 4D bilinear covariants there must be some redundancy in the five dimensional case. Indeed, the fact that we have a central pseudoscalar in the algebra provides this redundancy. For instance the pseudoscalar $\overline\psi{I}\psi=I\overline{\psi}\psi$ of the 5D spinor is just the scalar multiplied by the complex unit. As a result, it is enough for us to consider only the following bilinear covariants:
\begin{equation}\label{bilcov5}
\sigma=\bar{\psi}\psi,\;\;V^{A}=-i\bar{\psi}e^{A}\psi,\;\;M^{AB}=\frac{i}{2}\bar{\psi}[e^{A},e^{B }]\psi,
\end{equation}
where $\sigma$ is the scalar, $V^{A}$ the pseudovector and $M^{AB}$ the bivector. We have picked these particular covariants because they are all real. 
As we will explain later, we pick a matrix representation satisfying $\gamma_{\mu}=\tilde{e}_{\mu}=-ie_{4}e_{\mu}$ and $e_{4}=\gamma_{5}$ (with $\gamma_{0}$ and $\gamma_{5}$ in the Dirac representation), which is equivalent to picking $f=\frac{1}{2}(1+\tilde{e}_{0})\frac{1}{2}(1+i\tilde{e}_{1}\tilde{e}_{2})$ as the primitive idempotent generating the algebraic spinors (this also makes $\overline{\psi}=\psi^{\dagger}\gamma^{0}$ and hermitian adjoint coincide with Clifford conjugation). Hence, in terms of the gamma matrices the bilinear covariants are:
\begin{equation}\label{bilcovgamma}
\begin{gathered}
\sigma=\bar{\psi}\psi,\;\;V^{\mu}=\bar{\psi}\gamma^{5}\gamma^{\mu}\psi=K^{\mu},\;\;\\V^{5}=-i\bar{\psi}\gamma^{5}\psi=\omega,\;\;M^{\mu4}=\bar{\psi}\gamma^{\mu}\psi=J^{\mu},\;\; M^{\mu\nu}=\bar{\psi}[\gamma^{\mu},\gamma^{\nu}]\psi=S^{\mu\nu}.
\end{gathered}
\end{equation}
Lets note that the following equations hold:
\begin{subequations}
\begin{eqnarray}\label{fi1}
V_{A}V^{A}&=&-\sigma^{2}\\
M_{AB}M^{AB}&=&-V_{A}V^{A}=\sigma^{2}\\
M_{AB}V^{A}&=&0.\label{fi3}  
\end{eqnarray}\label{fi}
\end{subequations}
As we can see, the pseudoscalar is not an independent quantity any more, but the fifth component of the pseudovector $V^{A}$. Hence in addition to the removal of the six degrees of freedom in four dimensions, we can eliminate the pseudoscalar by a time-time rotation, amounting to a total elimination of seven degrees of freedom. Note that the order of these transformations is important. One possibility is to perform a time-time rotation making $\omega=0$, and then continue with the usual three boosts setting $J^{i}=0$ and the three rotations aligning $K^{i}$ with the $z$-axis. Because of \ref{fifi2}, performing the time-time rotation after the other six transformations, would create a non-vanishing space component for $J^{\mu}$, interfering with the work of the original boosts. In conclusion, we can set the spinor \ref{4ddof} to have $\omega=2\phi^{2}\sin{\beta}=0$, reducing it to the form:
\begin{equation}\label{5ddof}
\psi=\phi\begin{pmatrix}
		1\\
		0 \\
		1 \\
		0 \\
\end{pmatrix},
\end{equation}
which is in agreement with the previous work \cite{Vignolo}.

Since we are in a 5D spacetime, we could also wonder if we it would be possible to use the three boosts with respect to the extra time to remove degrees of freedom from the spinor \ref{4ddof}. This would seem paradoxical since there are at most only two degrees of freedom left to remove.  In order to perform such a boost, note that while $U^{A}$ will transform as a vector under the transformation, $J^{\mu}=M^{\mu4}$ will transform as the components of a rank two tensor. We have arrived at the spinor \ref{4ddof} by aligning the space vector $K^{j}$ to the $z$-axis, hence $K^{1}=K^{2}=0$, also $K^{0}=0$ (because $J^{i}=0$), which amounts to $V^{A}=(0,0,0,K^{3},\omega)$. Any boost involving directions $x$ and $y$ would ruin our alignment of $K^{i}$ with the $z$-axis, hence those are not free boosts for us to perform. However, we can perform an extra-time boost in the $z$-direction in order to set $V^{4}=0$. This will then reduce the form of the spinor \ref{4ddof} to \ref{5ddof}, which sets $\omega=0$. Note that this reduction would be equivalent to the one made using a time plane rotation.

For singular spinors the discussion is very similar. Let's observe that $\sigma=0$ is an  independent condition, but $\omega=0$ is not as strong as in four dimensions. If we assume $\sigma=0$ and $\omega=0$ then $V^{\mu}V_{\mu}=-V^{4}V^{4}=-\omega^{2}=0$, however we can perform a rotation in the time plane to make $V^{4}=\omega\neq{0}$, hence the case reduces to the case of regular spinors with $\beta=\pi/2$, (since $\sigma=2\phi^{2}\cos(\beta)=0$). In accordance with reference \cite{Vignolo} we arrive at the conclusion that every singular spinor can be written as:
\begin{equation}\label{4ddof}
\psi=\phi\begin{pmatrix}
		e^{i\pi/4}\\
		0 \\
		e^{-i\pi/4} \\
		0 \\
\end{pmatrix}.
\end{equation}

\subsection{Dirac equation in Minkowski spacetime $\mathbb{R}^{1,3}$ from $\mathbb{R}^{2,3}$ for neutral matter}

If we consider the four dimensional Minkowski spacetime $\mathbb{R}^{1,3}$.
The Dirac equation in the complexified Clifford algebra $\text{Cl}_{1,3}(\mathbb{C})$ is:
\begin{equation}
ie_{\mu}\partial^{\mu}\Psi-\frac{mc}{\hbar}\Psi=0,
\end{equation}
with $\Psi$ a Dirac spinor, $m$ the mass of the spinor field and $i$ the complex imaginary unit. In this section we shall consider the Minkowski 5D spacetime with an extra time-like dimension, $\mathbb{R}^{2,3}$ described in the previous section. An important thing to state about the coordinates is that, although $x_{0}$ and $x_{4}$ are time-like, all the coordinates have dimensions of length; hence $x_{0}=ct_{0}$ and $x_{4}=ct_{4}$. Since $x_{0}$ and $x_{4}$ are both time-like, we will call the plane $x_{0}-x_{4}$ the \emph{time-plane}.

Within the Induced Matter Theory (IMT) \cite{Wesson, STM, ParticleQM}, a way to treat mass in the 4D spacetime is to induce it from a 5D spacetime, as a property of the particle motion in the fifth direction. In our case, we shall apply this to the Dirac equation. In order to arrive to a massive 4D Dirac equation, we shall propose the massless 5D Dirac equation :
\begin{equation}\label{5DDiraceqn}
e_{B}\partial^{B}\Psi=0,
\end{equation}
which can be written in the following form:
\begin{equation}
e_{\mu}\partial^{\mu}\Psi+e_{4}\partial^{4}=0 \iff (e_{4}e_{\mu})\partial^{\mu}\Psi+(e_{4})^{2}\partial^{4}\Psi=0 \iff (e_{4}e_{\mu})\partial^{\mu}\Psi+\partial^{4}\Psi=0.
\end{equation}
Using the elements $\tilde{e_{\mu}}$ defined in the previous section and assuming $\partial^{4}\Psi=i\frac{p^{4}}{\hbar}\Psi=-\frac{mc}{\hbar}\Psi$, the equation is written:
\begin{equation}\label{InducedDirac}
i\tilde{e_{\mu}}\partial^{\mu}\Psi-\frac{mc}{\hbar}\Psi=0,
\end{equation}
and due to the fact that the elements $\tilde{e_{\mu}}$ obey the rules for the Clifford algebra $\text{Cl}_{1,3}$, we recover the familiar 4D Dirac equation.

Let's solve the Dirac equation \ref{InducedDirac}, subject to the condition $\partial^{i}\Psi=0 \ \forall{i\in\{1,2,3\}}$. This is, the spinor field is homogeneous in space, but may change in time. Then the equation is written:
\begin{equation}
i\tilde{e}_{0}\partial^{0}\Psi+\partial^{4}\Psi=e_{4}{e}_{0}\partial^{0}\Psi-\frac{mc}{\hbar}\Psi=0 \ \implies \ e_{4}{e}_{0}\partial^{0}\Psi=\frac{mc}{\hbar}\Psi.
\end{equation}
Multiplying by $e_{0}e_{4}$ on both sides we get:
\begin{equation}\label{PoincAlgRel}
\partial^{0}\Psi=\frac{mc}{\hbar}e_{0}{e}_{4}\Psi.
\end{equation}
Let's recall that $\partial^{0}$ is the generator of time-translations and $e_{0}{e}_{4}$ is a generator of a rotation in the plane $x_{0}-x_{4}$. Hence, this equation provides an equivalence between two transformations of the spinor field that we are to explore.
Since we have $\partial^{4}\Psi=-mc/\hbar\Psi$, it follows that $\Psi(x_{0},...,x_{4})=\Phi(x_{0},...,x_{3})\exp(-\frac{mc}{\hbar}x_{4})$. The 4D spinor solution $\Phi$, can be Fourier-expanded:
\begin{equation}\label{spinorexpansion}
\Phi(x_{0},...,x_{3})=\sum_{i=1}^{4}\int{dpA^{i}(p^{0},...,p^{3})\exp(-i\frac{p^{\mu}}{\hbar}x_{\mu})}{\bm{u_{i}}},
\end{equation}
with $\{\bm{u}_{1},\bm{u}_{2},\bm{u}_{3},\bm{u}_{4}\}$ a basis for the spinor vector space and $A^{i}:\mathbb{R}^{1,3}\rightarrow{\mathbb{C}}$, functions. The integral is taken on the surfaces $p^{\mu}p_{\mu}={m^{2}c^{2}} (\equiv p^{A}p_{A}=0)$. The fact that $\partial^{0}$ is the generator of translations in the (time) direction $x^{0}$, means that the operator defined by
\begin{equation}
\exp\big(\alpha\partial^{0}\big)=\sum_{n=0}^{\infty}\frac{\alpha^{n}(\partial^{0})^{n}}{n!},
\end{equation}
complies with
\begin{equation}
\exp\big(\alpha\partial^{0}\big)\Psi(x_{0},...,x_{4})=\Psi(x_{0}+\alpha,x_{1},...,x_{4}).
\end{equation}
With some care we will write
\begin{equation}\label{timetranslation}
\exp\big(x_{0}\partial^{0}\big)\Psi(0,x_{1}...,x_{4})=\Psi(x_{0},x_{1},...,x_{4}),
\end{equation}
where the left hand side must be interpreted as: taking the spinor $\Psi$ as a function of $x_{0}$, applying $\exp(\alpha\partial^{0})$ (with $\alpha$ an external parameter) and then substituting $\alpha$ by the value $x_{0}$ and $x_{0}=0$, in the initial state. This is important because the operator $\exp(x_{0}\partial^{0})$, where $x_{0}$ is a variable, is NOT the translation of magnitude $x_{0}$ in the direction $x_{0}$.

Since we are to work with rest solutions ($\partial^{k}\Psi=0$), $p^{k}=0$ holds for every $k$ in equation \ref{spinorexpansion}, and since the integral in the same equation is taken over $p^{A}p_{A}=0$ we get that ${(p^{0})}^2={m^{2}c^2}$, which implies $p^{0}=\pm{mc}$ and we obtain for the full spinor:
\begin{equation}\label{restspinorexpansion}
\Psi(x_{0},x_{4})=\sum_{i=1}^{4}\bigg[A^{i}_{0}\exp\bigg(-i\frac{mc}{\hbar}x_{0}-\frac{mc}{\hbar}x_{4}\bigg)+B^{i}_{0}\exp\bigg(i\frac{mc}{\hbar}x_{0}-\frac{mc}{\hbar}x_{4}\bigg)\bigg]{\bm{u_{i}}},
\end{equation}
and for $x_{0}=0$:
\begin{equation}\label{initialrestspinorexpansion}
\Psi(0,x_{4})=\sum_{i=1}^{4}\mathcal{A}^{i}_{0}\exp\Big(-\frac{mc}{\hbar}x_{4}\Big){\bm{u_{i}}},
\end{equation}
with $\mathcal{A}^{i}_{0}=A^{i}_{0}+B^{i}_{0}$, and all of these numbers are complex constants. By virtue of equation \ref{PoincAlgRel}, and due to the fact that $[\partial^{0},e_{0}e_{4}]=0$, we have:
\begin{equation}\label{timetranslation}
\exp\big(\alpha\partial^{0}\big)\Psi=\sum_{n=0}^{\infty}\frac{\alpha^{n}(\partial^{0})^{n}}{n!}\Psi=\sum_{n=0}^{\infty}\bigg(\frac{mc\alpha}{\hbar}\bigg)^{n}\frac{(e_{0}e_{4})^{n}}{n!}\Psi=\exp\bigg(\frac{mc\alpha}{\hbar}e_{0}e_{4}\bigg)\Psi.\\
\end{equation}
The quantity $\exp\big(\frac{mc\alpha}{\hbar}e_{0}e_{4}\big)\in{\text{Cl}_{2,3}}$ can be computed via the series expansion in equation \ref{timetranslation}. We know that $(ie_{0}e_{4})^{2}=1$, which implies
\begin{equation}
(ie_{0}e_{4})^{n}=
  \begin{cases}
    1 & \ \text{if $n$ is even}\\
    ie_{0}e_{4} & \ \text{if $n$ is odd}\\
  \end{cases}.
\end{equation}
Hence,we have
\begin{equation}
\exp\bigg(\frac{mc\alpha}{\hbar}e_{0}e_{4}\bigg)=\cosh\bigg(\frac{-imc\alpha}{\hbar}\bigg)1+\sinh\bigg(\frac{-imc\alpha}{\hbar}\bigg)ie_{0}e_{4}.
\end{equation}
Using the identities $\cosh(ix)=\cos(x)$ and $\sinh(ix)=i\sin(x)$, together with the fact that $\cosh$ is even and $\sinh$ is odd, we have
\begin{equation}
\exp\bigg(\frac{mc\alpha}{\hbar}e_{0}e_{4}\bigg)=\cos\bigg(\frac{mc\alpha}{\hbar}\bigg)1+\sin\bigg(\frac{mc\alpha}{\hbar}\bigg)e_{0}e_{4}.\\
\end{equation}
Combining this result with equation \ref{timetranslation} and recalling the definition $i\tilde{e}_{0}=e_{4}e_{0}$, we get for rest spinor solutions
\begin{equation}\label{spinorevolution}
\Psi(x_{0},x_{4})=\bigg[\cos\bigg(\frac{mcx_{0}}{\hbar}\bigg)1-i\sin\bigg(\frac{mcx_{0}}{\hbar}\bigg)\tilde{e}_{0}\bigg]\Psi(0,x_{4}).
\end{equation}
In this fashion we obtain the ``coordinate-time evolution'' for the field from the initial value, using an element on the Spin group of the 5D Minkowski space.

\subsection{Time evolution in coordinate space-time}

The advantage of the treatment here is that the ``time-evolution'' operator represents also a coordinate transformation and not only an internal transformation of the spinor. Since this operator is the exponential of a real multiple of $e_{4}e_{0}$, it belongs to the real Lie group $\text{Spin}(2,3)$ and hence the adjoint action will transform a vector $V^{A}e_{A}$ into another vector $V'^{A}e_{A}$, with $V'_{A}V'^{A}=V_{A}V^{A}$. Let's consider a 5-vector $V^{\mu}e_{\mu}+V^{4}e_{4}$ and see how it is transformed by the element $\exp\big(\frac{mc\alpha}{\hbar}e_{0}e_{4}\big)$ via the adjoint representation. We have
\begin{equation}
\begin{gathered}
\exp\bigg(\frac{mc\alpha}{\hbar}e_{0}e_{4}\bigg)(e_{i})\exp\bigg(\frac{-mc\alpha}{\hbar}e_{0}e_{4}\bigg)=\\\\
=\bigg[\cos\bigg(\frac{mc\alpha}{\hbar}\bigg)1+\sin\bigg(\frac{mc\alpha}{\hbar}\bigg)e_{0}e_{4}\bigg]e_{i}\bigg[\cos\bigg(\frac{mc\alpha}{\hbar}\bigg)1-\sin\bigg(\frac{mc\alpha}{\hbar}\bigg)e_{0}e_{4}\bigg]=\\\\
=\bigg[\cos\bigg(\frac{mc\alpha}{\hbar}\bigg)1+\sin\bigg(\frac{mc\alpha}{\hbar}\bigg)e_{0}e_{4}\bigg]\bigg[\cos\bigg(\frac{mc\alpha}{\hbar}\bigg)1-\sin\bigg(\frac{mc\alpha}{\hbar}\bigg)e_{0}e_{4}\bigg]e_{i}=e_{i}
\end{gathered}.
\end{equation}
Therefore the transformation in $\text{SO}(2,3)$ does not affect the three spatial components of a vector $V^{i}$. Computing analogously for $e_{0}$ and $e_{4}$, we obtain
\begin{equation}
e_{0}\mapsto\exp\bigg(\frac{mc\alpha}{\hbar}e_{0}e_{4}\bigg)(e_{0})\exp\bigg(\frac{-mc\alpha}{\hbar}e_{0}e_{4}\bigg)=\cos\bigg(\frac{2mc\alpha}{\hbar}\bigg)e_{0}-\sin\bigg(\frac{2mc\alpha}{\hbar}\bigg)e_{4}\\
\end{equation}
\\
\begin{equation}
e_{4}\mapsto\exp\bigg(\frac{mc\alpha}{\hbar}e_{0}e_{4}\bigg)(e_{4})\exp\bigg(\frac{-mc\alpha}{\hbar}e_{0}e_{4}\bigg)=\cos\bigg(\frac{2mc\alpha}{\hbar}\bigg)e_{4}+\sin\bigg(\frac{2mc\alpha}{\hbar}\bigg)e_{0}\\
\end{equation}.
Using this, a vector $V^{A}e_{A}$ transforms according to
\begin{equation}
\bigg[V^{0}\cos\bigg(\frac{2mc\alpha}{\hbar}\bigg)+V^{4}\sin\bigg(\frac{2mc\alpha}{\hbar}\bigg)\bigg]e_{0}+\bigg[V^{4}\cos\bigg(\frac{2mc\alpha}{\hbar}\bigg)-V^{0}\sin\bigg(\frac{2mc\alpha}{\hbar}\bigg)\bigg]e_{4}+V^{i}e_{i}.
\end{equation}
This transformation is a rotation of angle $2mc\alpha{/\hbar}$ on the plane of two times $x_{0}-x_{4}$. Note that if we take the projection of this 5-vector on the hyperplane orthogonal to the direction $x_{4}$ (the 4D space $\{x_{4}\}^{\perp}$), $V^{\mu}e_{\mu}$ and compute the adjoint action on it, and then project again on the same hyperplane we obtain
\begin{equation}
V^{\mu}e_{\mu}\mapsto{\cos\bigg(\frac{2mc\alpha}{\hbar}\bigg)V^{0}e_{0}+V^{i}e_{i}}
\end{equation}.
This is clearly not an isometry, since $V'^{\mu}V'_{\mu}-V^{\mu}V_{\mu}=\sin^{2}({2mc\alpha}/{\hbar})V^{0}V_{0}$, which in general is not zero. This is not a problem since the quantity that should be preserved by the $\text{Spin}(2,3)$ group is the 5D norm.

\subsection{The rest solution and the particle-antiparticle character}\label{restsolnofield}

Let us recall that in the usual CA $\text{Cl}_{1,3}$, the generators $e_{\mu}$ admit the Dirac matrix representation given by the gamma matrices. In particular the matrix $\gamma_{0}$ representing $e_{0}$ is written in the form
\begin{equation}
\gamma_{0}=
\begin{pmatrix}
		1 & 0 & 0 & 0\\
		0 & 1 & 0 & 0\\
		0 & 0 & -1 & 0\\
		0 & 0 & 0 & -1\\
\end{pmatrix},
\end{equation}
which clearly means that the basis of spinors $\{\bm{u}_{1},...,\bm{u}_4\}$ is a basis of eigenvectors for the matrix $\gamma_{0}$. The interpretation of the Dirac spinor as a particle or antiparticle for the \emph{rest solution} ($p^{i}=0$) is given precisely by the eigenvalue of $\gamma_{0}$, associated to the eigenstate in consideration. If it's $+1$, it is a pure particle state and if it is $-1$, it is a pure antiparticle state.

Now, the treatment given in this article is different. We have decided to add an extra time-like coordinate and to interpret the ordinary 4D Clifford algebra as a subalgebra of it. In our case the elements appearing in the Dirac equation are not $e_{\mu}$, but $\tilde{e}_{\mu}=-ie_{4}e_{\mu}$, and because of this fact, the particle/antiparticle interpretation of the rest solution should be given by the eigenvalues of $\tilde{e}_{0}$ instead of $e_{0}$ alone. As was stated before, seeing the 4D CA as this particular subalgebra generated by $\tilde{e}_{\mu}$, has the advantage that $ie_{\mu}$ lies in the Lie algebra of the Spin$(2,3)$ group and hence it is the generator of certain coordinate isometry.
Due to the facts stated above, and in order to work in a spinor basis of pure particle/antiparticle rest states, we need to change our matrix representation: instead of taking $e_{0}$ to be diagonal and equal to $\gamma_{0}$, we demand $\tilde{e}_{0}=-ie_{4}e_{0}=\gamma_{0}$. By consistency of the CA commutation relations, this implies also that $\tilde{e}_{\mu}=\gamma_{\mu}$, for every $\mu\in\{0,...,3\}$. Recall that $e_{4}=\gamma_{5}$, hence we have $\tilde{e}_{\mu}=\gamma_{\mu}=-i\gamma_{5}e_{\mu}$, and multiplying by $i\gamma_{5}$ on both sides, we have $e_{\mu}=i\gamma_{5}\gamma_{\mu}$. With this, the new representation for the generators $e_{A}$ is given by
\begin{equation}
e_{0}=
\begin{pmatrix}
		0 & 0 & -i & 0\\
		0 & 0 & 0 & -i\\
		i & 0 & 0 & 0\\
		0 & i & 0 & 0\\
\end{pmatrix}; \ \
e_{k}=
\begin{pmatrix}
		-i\sigma_{k} & 0 \\
		0 & i\sigma_{k} \\
\end{pmatrix}; \ \
e_{4}=
\begin{pmatrix}
		0 & 0 & 1 & 0\\
		0 & 0 & 0 & 1\\
		1 & 0 & 0 & 0\\
		0 & 1 & 0 & 0\\
\end{pmatrix},
\end{equation}
where the matrices $\sigma_{k}$ are the well known Pauli matrices (with $\sigma_{i}^{2}=1$). Since $\tilde{e}_{0}$ is diagonal, we have that the basis
\begin{equation}
\bm{u}_{1}=\begin{pmatrix}
		1 \\
		0 \\
		0 \\
		0 \\
\end{pmatrix}
;\quad
\bm{u}_{2}=\begin{pmatrix}
		0 \\
		1 \\
		0 \\
		0 \\
\end{pmatrix}
;\quad
\bm{u}_{3}=\begin{pmatrix}
		0 \\
		0 \\
		1 \\
		0 \\
\end{pmatrix}
;\quad
\bm{u}_{4}=\begin{pmatrix}
		0 \\
		0 \\
		0 \\
		1 \\
\end{pmatrix}
\end{equation}
is a basis of eigenstates of $\tilde{e}_{0}=-ie_{4}e_{0}$ with $\bm{u}_{1}$ and $\bm{u}_{2}$ having eigenvalue $+1$, and $\bm{u}_{3}$ and $\bm{u}_{4}$ having $-1$ as eigenvalue.

Let us use this fact for the rest solution \ref{spinorevolution}, with the integral expansion \ref{initialrestspinorexpansion}
\begin{equation}\label{initialreststate}
\Psi(0,x^{4})=\sum_{i=1}^{2}\mathcal{A}^{i}_{0}e^{-\frac{mc}{\hbar}x_{4}}{\bm{u_{i}}}+\sum_{j=3}^{4}\mathcal{A}^{j}_{0}e^{-\frac{mc}{\hbar}x_{4}}{\bm{u_{j}}}.
\end{equation}
In the following, when writing $\bm{u}_{i}$ we will be assuming $i\in\{1,2\}$ and when writing $\bm{u}_{j}$, $j\in\{3,4\}$. Applying the operator $\exp(\frac{-imcx^{0}}{\hbar}\tilde{e}_{0})$ on the $\bm{u}_{i}$ and $\bm{u}_{j}$ we have:
\begin{equation}
\begin{gathered}
\bigg[\cos\bigg(\frac{mcx_{0}}{\hbar}\bigg)1-i\sin\bigg(\frac{mcx_{0}}{\hbar}\bigg)\tilde{e}_{0}\bigg]\bm{u}_{i}=\bigg[\cos\bigg(\frac{mcx_{0}}{\hbar}\bigg)1-i\sin\bigg(\frac{mcx_{0}}{\hbar}\bigg)\bigg]\bm{u}_{i}=e^{-i\frac{mcx_{0}}{\hbar}}\bm{u}_{i}\ , \\
\bigg[\cos\bigg(\frac{mcx_{0}}{\hbar}\bigg)1-i\sin\bigg(\frac{mcx_{0}}{\hbar}\bigg)\tilde{e}_{0}\bigg]\bm{u}_{j}=\bigg[\cos\bigg(\frac{mcx_{0}}{\hbar}\bigg)1+i\sin\bigg(\frac{mcx_{0}}{\hbar}\bigg)\bigg]\bm{u}_{j}=e^{i\frac{mcx_{0}}{\hbar}}\bm{u}_{j}\ .
\end{gathered}
\end{equation}
Putting this back in equation \ref{spinorevolution}, and with \ref{initialreststate}, we have
\begin{equation}
\Psi(x^{0},x^{4})=\sum_{i=1}^{2}\mathcal{A}^{i}_{0}e^{-\frac{mc}{\hbar}(x_{4}+ix_{0})}{\bm{u_{i}}}+\sum_{j=3}^{4}\mathcal{A}^{j}_{0}e^{-\frac{mc}{\hbar}(x_{4}-ix_{0})}{\bm{u_{j}}}.
\end{equation}
Hence, a full particle rest solution will be characterized by $\mathcal{A}^{j}_{0}=0$, and a full antiparticle rest solution by $\mathcal{A}^{i}_{0}=0$, as usual. We can see that in both cases the solution oscillates in ordinary time, while it's damped in the extra-time.
Let us note, that we have solved only the rest case because it was sufficient for us and it simplified the calculus. However, one can build any solution in the series expansion \ref{spinorexpansion} with $p^{i}\neq{0}$ by just taking a representant $\bm{S}$ in the Spin group of the Lorentz transformation $\Lambda^{\mu}_{\nu}$ that takes $(mc,0,0,0)$ to certain $(p^{0},p^{1},p^{2},p^{3})$, and then transforming the spinors $\bm{u}_{i}$ into the states $\bm{u}'_{i}=\bm{S}\bm{u}_{i}$.

The possibility of a possitive mass associated to neutrinos and negative mass to antineutrinos was considered by Barut and Ziino in \cite{Barut}. In the article mentioned, they work under the hypothesis that neutrinos obey a Dirac equation with possitive mass and antineutrinos the same equation but with a negative mass term. Under this assumption a relationship between these two types of particles is obtained, accounting to an explanation of the parity violation for neutrinos and the non-existence of a right-handed neutrino.

On the contrary, in this article, we only consider a positive mass Dirac equation, which in the rest frame it reduces to the ordinary Dirac equation for particles, and to the negative mass Dirac equation for antiparticles. Since our treatment follows classical single Dirac equation, the particle/antiparticle character of the fermion is an instrinsic property only defined in the rest frame, which makes particles and antiparticles not absolutely separable, meaning that a boost in a given direction will in general mix positive eigenstates with negative eigenstates of $\tilde{e}_{0}$. Instead, the treatment provided in \cite{Barut} duplicates the fermion space, incorporating a mass operator which determines the particle/antiparticle quality, allowing a strong symmetry between chiral fermions and Dirac fermions.

Recall that the chiral decomposition of spinors is really succesfull in four dimensions because the projections $P_{R}=\frac{1}{2}(1+\gamma_{5})$ and $P_{L}=\frac{1}{2}(1-\gamma_{5})$ commute with every element of $\text{Spin}(1,3)$, hence neither the boosts nor the rotations mix right and left chiral spinors. We see that in five dimensions this spliting is lost since now we can perform boosts in the extra time $x_{4}$, mixing left and right chiral spinors. However, particles and antiparticles do not constitute such an elegant decomposition, since, as stated above, the boosts always mix them. We may think though, that particles and antiparticles wouldn't mix under extra time boosts, and that would have been the case if we had identified particles and antiparticles with eigenvectors of $e_{0}$, but since we considered $\tilde{e}_{0}$ instead, this is not the case.

The question of discrete symmetries in this theory is a very interesting point. In the work of Barut and Ziino, the fact that particles and antiparticles are parity eigenstates is central to the inference of other discrete transformation laws. If we imitate that principle in our formalism, the parity transformation would have to be $\psi\mapsto\tilde{e}_{0}\psi$, but this will only be the parity transformation for tilded vectors.

Another possibility is to consider the ordinary parity transformation $\psi\mapsto{e}_{0}\psi$, which on vectors reads:
\begin{equation}
e_{0}\mapsto{e_{0}}, \ \ e_{i}\mapsto-e_{i} \ \ \text{and} \ \ e_{4}\mapsto-e_{4}.
\end{equation}
We see that when restricted to the first four coordinates, it is the ordinary parity transformation. This parity transformation would produce an inversion in the sign of rest mass, since  $p^{4}=imc/\hbar$ would change to $-imc/\hbar$. This accounts for a parity violation in the Dirac equation which doesn't show up in four dimensions, but that is compatible with \cite{Barut}. A disadvantage of choosing this parity transformation is that we lose the identification of particles and antiparticles as parity eigenstates.

From the reasoning above, we see that some of the discrete transformations may have no straight forward definition in the five dimensional approach. The topics of parity, charge conjugation and time reversal are very interesting topics to study, in particular in relation with \cite{Barut}, where they obtain a covariant charge conjugation operator. Nonetheless, this treatment would be too extensive to be treated here, and we plan to explore it thoroughly in future work.

\subsection{The effective scale of the spinor field in the extra dimension}

Since $\partial^{4}{\Psi}=-mc/{\hbar}\Psi$ holds for any spinor solution, all massive spinors decay exponentially in the extra dimension. We see that if we assume that the extra coordinate doesn't extend to $-\infty$, then after some extra-time the field values get very small, the exponential damping constant being $mc/\hbar$, the inverse of the reduced Compton wavelength for the particle. Then for a given particle after a time of order $\hbar/mc^2=\lambdabar/c$ the field is $90\%$ smaller than initially. This would mean that even though we didn't propose the extra dimension to be bounded, the field would only extend over a finite region of it, of the order of the Compton wavelength (or $\lambdabar/c$ in units of time).

\subsection{Charged Dirac fields with an electromagnetic field}

The 5D massless Dirac equation has the global $U(1)$ symmetry given by $\Psi\mapsto{e^{i\alpha}\Psi}$, where $\alpha$ is a real number. QED introduces the electromagnetic field as the gauge field required to make this symmetry a local one. This is, the theory is invariant for $\Psi\mapsto{e^{i\alpha(x)}\Psi}$, with $\alpha(x)$ a function on the coordinates of the space-time. Usually $\alpha$ is a function of the four variables of space-time, but since now our spacetime has five coordinates, let's consider $\alpha=\alpha(x_{B})$. This induces the gauge covariant derivative: \[D^{B}=\partial^{B}+i\frac{q}{\hbar}A^{B} \ ,\] with the transformation for the spinor and gauge field $A^{B}$:
\begin{equation}
\begin{gathered}
\Psi\mapsto{e}^{i\alpha}{\Psi},\\
A^{B}\mapsto{A}^{B}-\frac{i\hbar}{q}\partial^{B}\alpha ,
\end{gathered}
\end{equation}
where $q$ is the charge of the field $\Psi$ in consideration. In this way we get an electromagnetic $5$-vector potential. In consequence, we have the massless Dirac equation
\begin{equation}
{e_{B}}D^{B}\Psi=0,
\end{equation}
with $D^{B}:=\partial^{B}+i\frac{q}{\hbar}A^{B}$. Proceeding as in the previous section we multiply on the left by $e_{4}$ and get
\begin{equation}
i{\tilde{e}_{\mu}}D^{\mu}\Psi+D^{4}\Psi=0,
\end{equation}
with $\tilde{e}_{\mu}=-ie_{4}e_{\mu}$, as in the previous sections. In order for the 4D Dirac equation to be fulfilled, we require $D^{4}\Psi=-\frac{mc}{\hbar}\Psi$. We see that the $4$-electromagnetic vector potential in the 4D induced CA is the tilded vector $A^{\mu}\tilde{e}_{\mu}$, since it is the quantity appearing in the Dirac equation. If we further require the solution to comply with $D^{i}\Psi=0$, then what is left is the equation
\begin{equation}
i{\tilde{e}_{0}}D^{0}\Psi-\frac{mc}{\hbar}\Psi=0.
\end{equation}
Multiplying by $-i\tilde{e}_{0}=-e_{4}e_{0}=e_{0}e_{4}$ on both sides, and using $D^{0}=\partial^{0}+i\frac{q}{\hbar}A^{0}$ we obtain
\begin{equation}\label{eqnwithemfield}
\partial^{0}\Psi=\bigg(\frac{mc}{\hbar}e_{0}e_{4}-i\frac{q}{\hbar}A^{0}\bigg)\Psi.
\end{equation}
Let's note that the condition we are asking for ($D^{i}\Psi=0$) doesn't make the field homogeneous. Hence now we have $\Psi=\Psi(x^{0},x^{1},x^{2},x^{3},x^{4})$.

In the previous section we went from equation \ref{PoincAlgRel} to \ref{timetranslation} very easily,  because $[\partial^{0},e_{0}e_{4}]=0$.
Here we have $\frac{mc}{\hbar}e_{0}e_{4}-i\frac{q}{\hbar}A^{0}$ in the right hand side of \ref{eqnwithemfield}, instead of just $\frac{mc}{\hbar}e_{0}e_{4}$. This leads us, unless $A^{0}$ doesn't depend on $x_{0}$, to the fact that $[\partial^{0},\frac{mc}{\hbar}e_{0}e_{4}-i\frac{q}{\hbar}A^{0}]\neq0$. Hence, the equation in the Lie algebra representation doesn't pass so easily to the Lie group representation.

\subsubsection{The case $\partial^{0}A^{0}=0$.}

In this case, we are in the conditions of neutral matter and by the same procedure we write:
\begin{equation}\label{solutionfield}
\Psi(x_{A})=\exp\bigg(\frac{mcx_{0}}{\hbar}e_{0}e_{4}-i\frac{q}{\hbar}A^{0}x_{0}\bigg)\Psi(0,x_{i},x_{4})
\end{equation}
In the previous section we interpreted the right hand side of equation \ref{spinorevolution} as an equation in the algebra $\bigwedge_{2}(\mathbb{R}^{2,3})$ which is the Lie algebra $so(2,3)$, however, we see now in equation \ref{eqnwithemfield} that the right hand member doesn't lie in $\bigwedge_{2}(\mathbb{R}^{2,3})$ but rather in $\bigwedge_{2}(\mathbb{R}^{2,3})\oplus{Z}(\text{Cl}(2,3))$, which is the Lie algebra of the Clifford-Lipschitz group $\Gamma_{2,3}$ (cf. section \ref{CLgroup}). This group contains all the elements that under the adjoint action on the CA transform vectors into vectors. Indeed the adjoint action of $\exp(-i\frac{q}{\hbar}A^{0}x_{0})$ on every element of the CA is the identity, since this element sits on the center of the CA. This can be put as follows: Additionally to the coordinate transformation that we have in the absence of electromagnetic field, there is also an internal transformation, which has no coordinate consequences, and that corresponds to the internal local gauge freedom of the theory. Note that this case contains every electrostatic potential $A^{0}(x_{1},x_{2},x_{3},x_{4})$. If we further ask for the cyclic condition to hold\cite{Wesson} (potentials not depending on the 5th coordinate), we have $A^{0}=A^{0}(x_{i})$.
Now, if we decide to use a basis of eigenvectors of $e_{0}e_{4}$ for the spinor, as in section \ref{restsolnofield}, we have that a $+i$ eigenvector solution is
\begin{equation}
\Psi(x^{0})=\exp\bigg(\frac{imcx_{0}}{\hbar}-i\frac{q}{\hbar}A^{0}x_{0}\bigg)\Psi_{0},
\end{equation}
and a $-i$ eigenvector solution 
\begin{equation}
\Psi(x^{0})=\exp\bigg(-\frac{imcx_{0}}{\hbar}-i\frac{q}{\hbar}A^{0}x_{0}\bigg)\Psi_{0}.
\end{equation}
This is quite different from what we obtained in section \ref{restsolnofield}, where the eigenvalue of $e_{0}e_{4}$ alone determined the particle/antiparticle state. Now this interpretations has to change, since an antiparticle state has an opposite charge and, as we see in the equations above, solutions associated to different eigenvalues of $e_{0}e_{4}$ have the same charge.
It happens that the interpretation of a particle/antiparticle in the charged case has not only to do with the coordinate rotation in the time-plane, but also with the local gauge transformation of the spinor in \ref{solutionfield}. In the neutral matter case, it was sufficient with changing the sense of a rotation in the time-plane, but now we also have to change the sense of ``rotation'' in the local gauge transformation, passing from $\exp(-i\frac{q}{\hbar}A^{0}x_{0})$ to $\exp(i\frac{q}{\hbar}A^{0}x_{0})$. Of course, the antiparticle covariant derivative is now different, with the opposite sign charge, and in consequence the Dirac equation satisfied is different. Note that when putting $q=0$, we recover everything in section \ref{restsolnofield}.

\subsubsection{The case $\partial^{0}A^{0}\neq{0}$.}

In this case we have: \[\bigg[\partial^{0},\frac{mc}{\hbar}e_{0}e_{4}-i\frac{q}{\hbar}A^{0}\bigg]=-i\frac{q}{\hbar}\partial^{0}(A^{0})\bm{1},\]
which makes a little harder to see the interpretation of the time evolution as a coordinate transformation. However, it is simple to solve directly the equation \ref{eqnwithemfield}, and then interprete the time evolution accordingly. Combining \ref{eqnwithemfield} with $D^{i}\Psi=0$ and $D^{4}\Psi=-mc/\hbar\Psi$, we have the solution
\begin{equation}
\Psi(x_{A})=\exp\bigg(\frac{mcx_{0}}{\hbar}e_{0}e_{4}-\frac{mcx_{4}}{\hbar}-i\frac{q}{\hbar}F(x_{A})\bigg)\Psi_{0},
\end{equation}
with $\partial^{B}{F}=A^{B}$. If we assume the solution to be separable, then $F$ is written as $F(x_{A})=F^{0}(x_{0})+F^{i}(x_{i})+F^{4}(x_{4})$ and consequently we can separate the dependence on $x_{0}$, from the rest of the coordinates
\begin{equation}
\Psi(x_{A})=\exp\bigg(\frac{mcx_{0}}{\hbar}e_{0}e_{4}-i\frac{q}{\hbar}F^{0}(x_{0})\bigg)\Psi_{0}(x_{i},x_{4}).
\end{equation}
Therefore, we recover the interpretation given in the previous case: the time evolution is a rotation in the time plane, together with an internal local gauge transformation, which now is not linear in time. And also, the antiparticle solution is obtained by changing the sense of rotation in the time-plane and in the internal transformation. The same observations about particle and antiparticles for the case $\partial^{0}A^{0}=0$ can be made here.

\subsection{Propagation in the extra dimension and 5D gauge fixing}

Since $D^{4}{\Psi}=(\partial^{4}+iq/{\hbar}A^{4})\Psi=-mc/{\hbar}\Psi$ holds for the charged spinors, then \[\partial^{4}\Psi=-\bigg(i\frac{q}{\hbar}A^{4}+\frac{mc}{\hbar}\bigg)\Psi , \] and in addition to the exponential decay in the case of neutral matter, there is also a propagation in the extra dimension. The only possibility that would forbid propagation in the extra time would be the 5th component of the vector potential being zero or imaginary. In order to have the same confinement as in the non-charged case we could ask for the gauge condition $A^{4}=0$, and in this case the extra dimension has the same effective extension for both the charged and the neutral case. Let's note that this condition, would imply that the theory is invariant only under the local gauge transformation  $\exp(i\alpha)$, with $\partial^{4}\alpha=0$. This is, the field have the same phase along any line $(x_{0},x_{i},x_{4}=\lambda)$, with $\lambda$ ranging over $\mathbb{R}$. Note also that if in addition we ask for $\partial^{4}A^{\mu}=0 \ \forall \mu$ then $F^{AB}F_{AB}=F^{\mu\nu}F_{\mu\nu}$ and we recover the classical QED Lagrangian.

\section{On ghosts and tachyons in this two-times physical theory}

The typical problems in theories with more than one timelike dimension are the existence of tachyons and the existence of ghost fields (negative norm states). This two issues makes it difficult to quantize the theory\cite{Bars}.

As exposed in \cite{2times}, the momentum in the extra dimension is related to the mass of the particle, hence tachyons are not a problem since there's no superluminical propagation in the 4D spacetime.

Regarding ghost fields, if we assume the following Lagrangian density:
\begin{equation}\label{lag1}
\mathcal{L}=\Psi^{\dagger}\gamma_{0}e_{A}\partial^{A}\Psi-\partial^{A}\Psi^{\dagger}\gamma_{0}e_{A}\Psi \ ,
\end{equation}

then the corresponding 5-current associated to the phase shift symmetry is given by:
\begin{equation}
J_{A}=i\Psi^{\dagger}\gamma_{0}e_{A}\Psi ,
\end{equation}
and if we look at the 0th component, we get $J_{0}=i\Psi^{\dagger}\gamma_{0}e_{0}\Psi=\Psi^{\dagger}\gamma_{5}\Psi $, which is not positive definite. Indeed, none of the components $J_{A}$ equals the probability density $\Psi^{\dagger}\Psi$.

However, recall that we have defined the 4D induced CA, as the one having generators $\{\tilde{e}_{\mu}\}$. In this context, the same equations of motion can be obtained from the following Lagrangian density:
\begin{equation}\label{lag2}
\mathcal{L}=\frac{\Psi^{\dagger}\gamma_{0}\tilde{e}_{\mu}\partial^{\mu}\Psi-\partial^{\mu}\Psi^{\dagger}\gamma_{0}\tilde{e}_{\mu}\Psi}{2} +\frac{i}{2} (\Psi^{\dagger}\gamma_{0}\partial^{4}\Psi -\partial^{4}\Psi^{\dagger}\gamma_{0}\Psi))\ ,
\end{equation}
which, after imposing $\partial^{4}\Psi=-\frac{mc}{\hbar}\Psi$ is written as:
\begin{equation}
\mathcal{L}=\frac{\Psi^{\dagger}\gamma_{0}\tilde{e}_{\mu}\partial^{\mu}\Psi-\partial^{\mu}\Psi^{\dagger}\gamma_{0}\tilde{e}_{\mu}\Psi}{2}+im\Psi^{\dagger}\gamma_{0}\Psi\ .
\end{equation}

This is the usual Dirac Lagrangian, since $\tilde{e}_{\mu}=\gamma_{\mu}$, and hence we have the usual conserved 4-current:
\begin{equation}
\tilde{J}_{\mu}=\Psi^{\dagger}\gamma_{0}\gamma_{\mu}\Psi=\Psi^{\dagger}\gamma_{0}\tilde{e}_{\mu}\Psi.
\end{equation}
In this way, the problem of ghosts is solved by considering $\mathbf{\tilde{J}}$ as the physical quantity, and not $\mathbf{J}$. Of course, $J_{A}$ is a 5-vector, while $\tilde{J}_{\mu}$ is a 4-vector. This means that when performing a Lorentz transformation involving the extra coordinate, in general $\tilde{J}_{\mu}$ won't obey a vector transformation rule, while $J_{A}$ will. Indeed, using this tilded current, the Dirac theory is completely unchanged.

Note that although $\tilde{J}^{\mu}$ is not obtained as the vector bilinear covariant of the field, it is a projection of the 3-vector bilinear covariant of the 5D CA (since each $\tilde{e}_{\mu}$ is a 3-vector), hence, the information is present in a bilinear covariant of the theory. This change from the, let's say, ``vector current'' to this ``3-vector current'' is related to the equivalence of Lagrangians (\ref{lag1}) and (\ref{lag2}) once we impose the constraint $\partial^{4}\Psi=-\frac{mc}{\hbar}\Psi$. This also implies that the 5 current $J_{A}$ is always conserved, while $\tilde{J}_{\mu}$ is only conserved once we introduce the aforementioned constraint. 

\section{Conclusions and prospects}

In this article we have obtained the 4D Dirac equation for spinors from a 5D massless equation. In order to do so, we have exploited the fact that the Dirac theory naturally comes equiped with an extra timelike dimension. Hence, following the principles of induced matter theory \cite{STM,Wesson} we have studied the Dirac equation in a 5D spacetime of signature $2,3$. Analyzing the Dirac equation for neutral matter (or absence of EM fields) and the Clifford algebra structure of the 5D space-time, we have provided an interpretation for a particle/antiparticle states, as eigenstates of the generators of rotations in the time-plane. In the charged case (with the presence of an EM field), by performing a similar analysis we have seen that in addition to the rotation in the time-plane, there is an internal gauge transformation in two opposite directions that separates particles from antiparticles. Additionally, we have observed that although the extra dimension was not proposed to be bounded, the effective extension of the field over it is of the order of the Compton wavelength of the particle.

In the absence of EM fields, the interpretation of particles and antiparticles given in this article could be analyzed by studying high order representations of the group $\text{SO}(2,3)$.  In just the same way spin up and down are the two states associated to the $s=1/2$ representation of the $\text{SO}(3)$ group, the particle and antiparticle states could be the two states of a particular representation of the group (or some subgroup of) $\text{SO}(2,3)$. If the extra dimension is a physical one and representation theory predicts more states of matter (other than matter/antimatter), it should be possible to try to observe them in experiments.

The relation of our work with \cite{Barut} is still to be further explored, particularly in relation to C, P, T and CPT symmetries. An interesting proposal would be to combine our analysis with the formalism of positive mass particles / negative mass antiparticles.

The main success of the Kaluza-Klein theory lies on its unification of Einstein field equations and Maxwell electromagnetism. It also introduces the Lorentz force on a test particle in the geodesic equations of motion. Due to these facts an interesting approach to induce an EM field in the Dirac theory (alternative to the used above) would be given by working with the original Kaluza-Klein metric \cite{Wesson}. In this way, we would have a non trivial 5D manifold and by working on the Clifford and spinor bundles we could try to obtain QED in absence of gravity. This is something we would like to do in the future since it would also allows us to work with gravitational fields.

\section*{Acknowledgements}

\noindent The authors acknowledge CONICET, Argentina (PIP 11220150100072CO) and UNMdP (EXA852/18), for financial support. M.R.A.A. thanks professor Rold\~{a}o da Rocha for reading the paper and for the references suggested regarding SO$(2,3)$.

\end{document}